\documentclass{amsart}
\usepackage{amssymb,amsmath,amsthm,amsfonts,epsfig}
\DeclareMathOperator{\rt}{root}

\begin{document}
\title[Taking Trees Seriously]{Mathematical Models and Biological Meaning: Taking Trees Seriously}
\date{July 10, 2008}
\author{Jeremy L.\ Martin}
\address{Department of Mathematics, University of Kansas, Lawrence, KS 66045, USA}
\email{jmartin@math.ku.edu}
\author{E. O. Wiley}
\address{Biodiversity Research Center, University of Kansas, Lawrence, KS 66045 USA, and
Department of Ecology and Evolutionary Biology, University of Kansas, Lawrence, KS 66045 USA}
\email{ewiley@ku.edu}
\maketitle
\begin{abstract}
We compare three basic kinds of discrete mathematical models used to portray phylogenetic relationships among species 
and higher taxa: phylogenetic trees, Hennig trees and Nelson cladograms.  All three models are trees, as that term is 
commonly used in mathematics; the difference between them lies in the biological interpretation of their vertices and 
edges.  Phylogenetic trees and Hennig trees carry exactly the same information, and translation between these two kinds 
of trees can be accomplished by a simple algorithm.  On the other hand, evolutionary concepts such as monophyly are 
represented as different mathematical substructures are represented differently in the two models.  For each 
phylogenetic or Hennig tree, there is a Nelson cladogram carrying the same information, but the requirement that all 
taxa be represented by leaves necessarily makes the representation less efficient.  Moreover, we claim that it is 
necessary to give some interpretation to the edges and internal vertices of a Nelson cladogram in order to make it 
useful as a biological model.  One possibility is to interpret internal vertices as sets of characters and the edges as 
statements of inclusion; however, this interpretation carries little more than incomplete phenetic information.  We 
assert that from the standpoint of phylogenetics, one is forced to regard each internal vertex of a Nelson cladogram as 
an actual (albeit unsampled) species simply to justify the use of synapomorphies rather than symplesiomorphies.
\end{abstract}

\section*{Introduction}
 
In the Willi Hennig Memorial Symposium, held in 1977 and published in Systematic Zoology in 1979, David Hull expressed 
the concern that ``uncertainty over what it is that cladograms are supposed to depict and how they are supposed to 
depict it has been one of the chief sources of confusion in the controversy over cladism'' (Hull, 1979:420). Early 
disagreements concerning the differences between cladograms and phylogenetic trees (Cracraft, 1974; Harper, 1976; 
Platnick, 1977; Wiley, 1979a, b, 1981a) were largely generated by such differences.  The purpose of this article is to 
compare three commonly used tree models of phylogenetic relationships, namely Hennig trees, phylogenetic trees, and 
cladograms sensu Nelson (1979), using the mathematical techniques of graph theory.  We assert that it is imperative to 
understand the mathematical relationships between these three kinds of models in order to make meaningful biological 
statements; that is, to reduce the confusion and ``uncertainty'' observed by Hull thirty years ago.

The vertices of a Hennig tree represent taxa (sampled or unsampled), while its edges model ancestry relationships. By 
contrast, in a phylogenetic tree, taxa are modeled by edges, while vertices correspond to speciation events.  These two 
models are isomorphic (as that term is used in mathematics) but not equal: that is, they carry exactly the same 
information about ancestry, but encoded in two different ways.  To make this explicit, we give a simple algorithm that 
constructs a unique Hennig tree for every phylogenetic tree and vice versa, and we explain how to translate key 
phylogenetic concepts (such as monophyletic groups) between these two mathematical models.

In a Nelson cladogram, sampled taxa are represented by leaf vertices; the edges and internal vertices ``illustrat[e] an 
unspecified relationship'' between the taxa (Nelson, 1979 ms).  We assert that cladograms in this sense are not useful 
as phylogenetic models without some phylogenetic interpretation for the edges and internal vertices.  The construction 
of a Nelson cladogram from character data (Nelson and Platnick, 1981), whether by parsimony or some other method, 
inherently includes an interpretation of each internal vertex as a set of apomorphies (evolutionary innovations) passed 
on all vertices descending (in the mathematical sense) from that vertex in unmodified or modified form.  However, even 
that interpretation carries little more than phenetic meaning (albeit a special similarity, not overall similarity).  
In order for the cladogram to have any phylogenetic significance, we claim that the vertices must be interpreted as 
hypothetical ancestral species, so that the entire cladogram represents a related group of species and to justify the 
use of apomorphies rather than plesiomorphies.  We also assert that the requirement that the internal vertices cannot 
represent taxa risks making the cladogram model less efficient or even less accurate.

\section*{Some Basic Graph Theory}

Mathematically speaking, all of the diagrams we shall consider are \emph{graphs}: they are finite structures built out 
of \emph{vertices} (sometimes called nodes) and \emph{edges}, in which each edge connects two vertices (see West, 1999) 
for background.  A graph is usually represented by drawing the vertices as dots and the edges as line segments.  
Frequently, the vertices and/or edges are labeled with names, numbers, or other data.  Graphs provide a simple and 
powerful tool to model and study phylogenetic and synapomorphic relationships between taxa (and many other structures); 
however, one must be very careful to keep track of what the individual vertices and edges are supposed to mean, 
particularly when (as here) there is more than one way to represent the same biological data in a graph.  Before 
proceeding, we mention a few basic facts and terms from graph theory, so as to have a unified mathematical language with 
which to work.  (We will introduce more technical material later, as needed.)

We will primarily be concerned with graphs that are \emph{trees}.  Mathematically, a tree is a graph $T$ containing no 
closed loops; intuitively, if you walk along the edges from vertex to vertex, the only way to return to your starting 
point is to retrace your steps.  If we designate one vertex $r$ as the \emph{root} of $T$, then every edge connects a 
vertex $x$ that is closer to $r$ with a vertex $y$ that is further away.  In this case, we say that $x$ is the 
\emph{parent} of $y$, and it is often convenient to regard the edge between them as a directed edge (or \emph{arc}) 
pointing from $x$ to $y$, represented by the symbol $x\to y$.  Every vertex in a tree has a unique parent, except for 
the root, which has no parent.  An immediate consequence is the useful fact that every tree with $n$ edges has $n+1$ 
vertices, and vice versa.  (Of course, several different vertices may share a common parent.)

The \emph{ancestors} of a vertex are its parent, its parent's parent, its parent's parent's parent, and so on.  
Equivalently, we might say that an edge $x\to y$ is an ancestor of another edge $a\to b$ if $y$ is equal to, or an 
ancestor of, $a$.  A \emph{lineage} (or \emph{ancestral lineage}) of a vertex $x$ is the complete list of vertices that 
are ancestors of $x$ and are descendants of, or equal to, some other vertex $y$.  If $y=\rt(T)$, then this 
list is called the \emph{total lineage} of $x$.  It is important to note that the choice of a root vertex, together with 
the topology of the tree, completely determines all ancestry relationships.

A \emph{subtree} of a tree $T$ is a tree $U$ all of whose vertices and edges are vertices and edges of $T$ as well.  
This is equivalent to saying that $U$ can be formed by removing some vertices and edges from $T$.  If in addition $T$ is 
a rooted tree, then $U$ inherits its ``ancestor-of'' relation from $T$ as well.  A \emph{proper subtree} of a rooted 
tree is a subtree that consists of a vertex and all its descendants.  A proper subtree is uniquely determined by its 
root vertex, so there are exactly as many proper subtrees of T as there are vertices.

Trees are well suited for modeling phylogenetic relationships between species or taxa, in which each species or taxon 
has a unique parent.  Uniqueness is vital; a tree cannot model, e.g., tokogenetic relationships in a sexually 
reproducing species (where each organism has two parents).

\section*{Phylogenetic Trees}

By the term \emph{phylogenetic tree}, we mean a tree that models (hypothesized) phylogenetic relationship among taxa by 
depicting taxa by edges, and speciation events by vertices.  For instance, in the phylogenetic tree in Fig.~1a, the 
terminal edges, labeled $A$, $B$, and $C$, represent named taxa; that is, large groups of individual organisms 
represented by sampled specimens.  The internal edges, labeled $y$ and $z$, represent ancestral lineages needed to 
account for the terminal taxa under the paradigm of descent with modification.  The vertices represent speciation 
events, in which the edge below the vertex is the common ancestor and the edges above it are descendants.  
Mathematically, the edge $y$ is the youngest common ancestor of the edges $B$ and $C$.  Biologically, moving up the tree 
represents moving forward in time, so the edge y represents a lineage of common ancestors of the sampled taxa $B$ and 
$C$, occurring before the speciation event that distinguishes $B$ and $C$ and after any previous speciation events.  
Thus the total lineage of a species (or, more properly, a hypothesis of its lineage) is represented by a chain of edges 
starting with the species itself and moving down the tree towards the root vertex, which necessarily has only one edge 
emanating from it---representing the common ancestor of all sampled taxa.

We frequently refer to the internal edges as ``hypothetical'' ancestors. However, under the paradigm of evolution, there 
is nothing more hypothetical about these edges than there are about the named taxa represented by specimens. Under the 
evolutionary paradigm, the extent to which we treat named taxa ($A$, $B$, $C$) as real entities of descent with 
modification is the extent to which we treat internal lines as symbolizing real ancestors. They are not 
``hypothetical''; they are simply unsampled.

In Fig.~1b, we have added more information to the phylogenetic tree.  Each numbered black rectangle represents an 
evolutionary character hypothesized to be fixed (sensu Wiens and Servedio, 2000) somewhere in the lineage represented by 
the edge to which the rectangle is attached.  (The placement of the rectangle within an edge does not matter; for 
example, the tree in Fig.~1b does \emph{not} assert that apomorphies 3, 4, and 5 became fixed at the same time just 
because they are shown at the same height on the page.  Moreover, one cannot draw inferences about when characters 
originated; for example, it is possible that character 2 originated in lineage $z$ before character 1, but went extinct 
in other lineages (such as $A$) and became fixed only in the common ancestor $y$ of $B$ and $C$.)

\section*{Hennig Trees And Phylogenetic Trees}

Hennig (1966) used the symbology of Gregg (1954), which Gregg derived from Woodger.  In a Hennig tree, taxa are 
represented by vertices, not by edges.  An edge of a Hennig tree does not represent a lineage or anything else occurring 
in nature: rather, it represents a relationship among two vertices, or more empirically, the hypothesis of a 
relationship.  Specifically, an edge between a parent vertex X and a child vertex Y represents the hypothesis that the X 
is an ancestor of Y.

Fig.~2 is redrawn from Hennig (1966) and portrays the relationships among samples of an evolving clade in two ways. The 
left-hand side of Fig.~2 portrays a phylogenetic tree with sampled populations of lineages (B1, B2, etc.) represented by 
edges (species to Hennig), with sampled populations placed in time with circles.  Vertices represent speciation events.

The right-hand side of Fig.~2 shows the Hennig tree (in our sense of the term) corresponding to the phylogenetic tree on 
the left-hand side.  Here the taxa are represented by vertices (ignoring the distinctions indicated by the numeric part 
of the labels).  The edges represent phylogenetic, not phenetic relationships between these species; Hennig makes this 
clear in a number of diagrams (Hennig, 1966, Figs.~4, 6, 14, 15) and in his text.

Indeed, we will prove mathematically that Hennig trees and phylogenetic trees carry the same information, albeit encoded 
in different ways.  We start by setting up some notation.

Let $T$ be a tree with root vertex $r$.  Recall that specifying a root for a tree determines its ``parent'' and 
``ancestor'' relations completely.  If $x$ is the parent of $y$, we will denote the edge joining them by the symbol 
$x\to y$ (in keeping with the convention that edges point from parents to children).  Alternately, we will write $x > y$ 
to indicate that vertex $x$ is an ancestor of vertex $y$.

It is a standard fact that for every set $X$ of vertices in $T$, there is a unique vertex $y$ (which may or may not 
belong to $X$) with the following two properties: first, $y\geq x$ for every $x$ in $X$, and second, if $z$ is any other 
vertex such that $z\geq x$ for every $x$ in $X$, then $z > y$.  The first of these conditions says that $y$ is a common 
ancestor of the vertices in $X$; the second condition says that it is the \emph{youngest} common ancestor.

Finally, we call $T$ a \emph{planted tree} if its root $r$ has only one child.  (``Planted'' is a more restrictive 
condition than ``rooted''; every planted tree is necessarily rooted, but not vice versa.)

We now describe an equivalence between two different kinds of labeled trees.  Let $n$ be any positive integer, and let 
$T$ be a rooted tree with $n$ vertices, labeled $1, 2, \dots, n$. (Any of these may be the root of $T$.)  Construct a 
tree $U$ from $T$ according to the following algorithm.

\vskip10bp
\textbf{Algorithm A}

\begin{enumerate}
\item Create a new root vertex, labeled 0, and create a new edge $0\to r$, where $r=\rt(T)$.
\item Label each edge $v\to w$ of this tree with the number $w$.
\item Erase the labels of the vertices.
\end{enumerate}

An example of the construction of $U$ from $T$ is shown in Fig.~3.  (The vertex labels are shown in blue, and the edge 
labels in red.)  Note that $U$ has $n+1$ vertices, hence $n$ edges, which are labeled $1, 2, \dots, n$.  A consequence 
of the construction is that $U$ is always a planted tree, because its root (from which the label 0 was erased) has 
exactly one child, namely, $r = \rt(T)$.

We can reconstruct $T$ from $U$ by reversing Algorithm~A.  Specifically, suppose that $U$ is any planted tree with $n$ 
edges, labeled $1, 2, \dots, n$.  Note that $U$ must have exactly $n+1$ vertices.  Let $r$ be the root vertex, and let 
$s$ be its unique child.  Now, construct a tree $T$ from $U$ as follows:

\vskip10bp
\textbf{Algorithm B}

\begin{enumerate}
\item Label each non-root vertex of $U$ by the label of its parent edge, and assign the label 0 to vertex $r$.
\item Erase all labels on the edges.
\item Delete vertex $r$ and edge $r\to s$, and designate $s$ as the root of the resulting tree.
\end{enumerate}

These steps are exactly the reverse of those of Algorithm~A; for an illustration, see Fig.~3.  (It is worth mentioning 
that the algorithms work the same way whether or not the input tree has polytomies (vertices with more than two 
children.)  The algorithms establish the following mathematical fact.

\vskip10bp
\textbf{Theorem 1}

There is a one-to-one correspondence between the following two sets:
\begin{itemize}
\item The set of all rooted trees $T$ on $n$ vertices labeled $1, 2, \dots, n$; and
\item The set of all planted trees $U$ on $n+1$ vertices, with edges labeled $1, 2, \dots, n$.
\end{itemize}

Because the correspondence is one-to-one, the rooted tree $T$ contains exactly the same information as its planted 
counterpart $U$.  However, one must be careful when translating between $T$ and $U$.  For example, there is not a 
one-to-one correspondence between subtrees of $T$ and subtrees of $U$.  Indeed, if $E$ is the set of edges of an 
arbitrary subtree of $U$, then the corresponding set of vertices of $T$ will not form a subtree unless $E$ is planted.  
For example, if $T$ and $U$ are as in Fig.~4(a,b), then edges 4, 5, 8, 9 form a proper subtree of $U$, but vertices 4, 
5, 8, 9 do not form a subtree of $T$.  On the other hand, vertices 2, 4, 5, 8, 9 do form a proper subtree of $T$, and 
the corresponding edges form a \emph{planted} proper subtree of $U$; see Fig.~4(c,d).

Indeed, it follows from Algorithms A and B that there is a one-to-one correspondence between proper subtrees of $T$ and 
planted proper subtrees of $U$.  Similarly, there is a one-to-one correspondence between subtrees of $T$ (not 
necessarily proper) and planted subtrees of $U$ (again, not necessarily proper).

Additional biological information associated with a phylogenetic or Hennig tree can be translated via this algorithm.  
For instance, the character data represented by edge labels in a phylogenetic tree (as in Fig.~2b) can be represented by 
vertex labels in the corresponding Hennig tree.

\section*{Two Concepts Of Monophyly Circumscription}

While Hennig trees and phylogenetic trees carry the same basic information about taxa and ancestry, they represent this 
information in different ways.  Therefore, it should not be surprising that biological concepts such as monophyly are 
modeled by different mathematical substructures in the two kinds of trees.  Hennig's (1966:206-209) discussion of 
monophyly admits only one definition of this term; a monophyletic group is a group that includes all descendants of a 
common ancestral species. Although not mentioned in this section, Hennig (1966:71) makes it clear that the ancestral 
species is also a member of the group (and, indeed is logically equivalent to all descendant members of the group). 
However, others have held that there are different ways to circumscribe monophyletic groups (de Queiroz and Gauthier, 
1990, 1992, 1994; de Queiroz, 2007) and this difference has been codified into formal rules that distinguish three kinds 
of clade recognition. The first two concern Hennig trees and phylogenetic trees (we will consider the third later).

Definition 1: ``A node-based clade is a clade originating from a particular node on a phylogenetic tree, where the node 
represents a lineage at the instant of a splitting event.'' (The PhyloCode, Note 2.1.4)

Definition 2. ``A branch-based clade is a clade originating from a particular branch (internode) on a phylogenetic tree, 
where the branch represents a lineage between two splitting events.'' (loc. cit.)

We suggest that the distinction between node-based and branch-based concepts of monophyly arises from confusion between 
the two types of trees we have discussed.  Indeed, adopting Hennig's (1966:71) usage of ``monophyly'', it becomes 
evident that a monophyletic group with common ancestor $A$ is represented in a Hennig tree $T$ by the proper subtree 
\emph{rooted} at the \emph{vertex} corresponding to $A$, and in a phylogenetic tree $U$ by the proper subtree 
\emph{planted} at the \emph{edge} corresponding to $A$.  (Recall that the proper subtrees of $T$ are in bijection with 
the planted proper subtrees of $U$.)  To rephrase this observation, the correct mathematical representation of monophyly 
can be found either by applying Definition~1 to a Hennig tree, or by applying Definition~2 to a phylogenetic tree.

It is worth examining what happens if we apply Definitions~1 and~2 to the wrong kinds of trees.  First, a ``node-based 
clade'' of a phylogenetic tree---speaking mathematically, a proper but non-planted subtree of a phylogenetic tree---does 
not correspond to a monophyletic group of taxa.  Returning to the phylogenetic tree $U$ shown in Fig.~3d, the 
non-planted subtree highlighted in Fig.~4(b) is actually polyphyletic, not monophyletic; every edge in $U$ represents a 
taxon descended from taxon 2, which does not belong to the subtree.  (That this set of taxa is polyphyletic is perhaps 
clearer upon examining the corresponding vertices in the Hennig tree; see Fig.~4(a).)  This matches the definition of 
``crown clade''.  Second, a planted subtree in a Hennig tree is not monophyletic but paraphyletic, because it includes 
only one child of its root vertex.  If we are careful only to use the term ``node-based clade'' when working with Hennig 
trees, and ``branch-based clade'' when working with phylogenetic trees, then the two terms become synonymous; the 
difference lies only in the representation and has no biological significance.

\section*{Nelson Cladograms}

Camin and Sokal (1965:312) coined the term ``cladogram'' to refer to a diagram that depicts the branching of a 
phylogenetic tree ``without respect to rates of divergence''. Gary Nelson (unpublished but widely cited manuscript, 
1979) had a much broader concept of cladograms and trees: ``A cladogram may, therefore, be defined as a dendritic 
structure illustrating an unspecified relationship between certain specified terms that in the context of systematics 
represent taxa,'' while ``a tree may be defined as a dendritic structure having one or more general as well as unique 
components (or combination of components).'' Nelson's ``dendritic structure'' must mean ``tree'' in the mathematical 
sense (that is, an acyclic connected graph), as noted by Hendy and Penny (1984).  Thus a cladogram sensu Nelson and a 
tree sensu Nelson differ in their biological interpretations, not their mathematical structure.  In Nelson's definition 
of a cladogram, the edges and internal vertices are left undefined.  On the other hand, it is necessary to assign 
\emph{some} biological meaning to the edges and internal vertices; otherwise, it is meaningless to assert that any 
particular cladogram is the ``right'' one for a given set of sampled taxa and character information.  Biologists 
(naturally enough) expect the internal structure of a cladogram to say something specific about evolutionary 
relationships (real or hypothesized) among species.

The simplest way to interpret an internal vertex $x$ of a Nelson cladogram is as the set of characters common to the 
species represented by leaf vertices descended from $x$. However, under this interpretation, the cladogram becomes 
nothing more than a phenogram.  We do not see a way to assign phylogenetic meaning to a cladogram without interpreting 
the internal vertices as ancestral species---which amounts to interpreting the cladogram as a Hennig tree.

What is the right way to translate between Hennig trees and Nelson cladograms?  Given a Hennig tree $H$ (which, as we 
now know, contains the same information as a phylogenetic tree), it is possible to construct a ``quasi-cladogram'' $Q$ 
containing the same information, but in which all taxa corresponding to vertices in $H$ are now represented by leaf 
vertices.  The tree $Q$ will have the property that each internal vertex is the parent of exactly one leaf.

\vskip10bp
\textbf{Algorithm C}

\begin{enumerate}
\item Find a vertex of $H$ with label $x$.  Introduce a new leaf adjacent to this vertex.  Label the new leaf $x$ 
and delete the label on the old vertex.
\item Repeat until $H$ has no labeled internal vertices.  Call the resulting tree $Q$.
\end{enumerate}

This construction is reversible via the following algorithm, which recovers $H$ from $Q$.

\vskip10bp
\textbf{Algorithm D}

\begin{enumerate}
\item Find an unlabeled vertex of $Q$.  It will have exactly one child that is a leaf. Move the label on that leaf to 
its parent, and delete the leaf.
\item Repeat until $Q$ has no labeled internal vertices.  Call the resulting tree $H$.
\end{enumerate}

\vskip10bp
\textbf{Theorem 2}

There is a one-to-one correspondence between the following two sets:
\begin{itemize}
\item The set of all rooted trees $H$ with vertices $1, 2, \dots, n$; and
\item The set of all rooted trees $Q$ with leaves $1, 2, \dots, n$, such that each internal vertex is the parent of 
exactly one leaf.
\end{itemize}

For an example of a pair of trees $H$ and $Q$ related in this way, see Fig.~5.  (Of course, not every cladogram arises 
in this way.)  It is worthwhile to examine the meaning of the quasi-cladogram $Q$.  For each leaf of $H$, its parent in 
$Q$ is an internal vertex with only one child; these are the vertices indicated by hollow circles in Fig.~5.  It is 
tempting, but not quite accurate, to remove those vertices (in mathematical terms, to contract one edge incident to each 
such vertex) so as to replace $Q$ with the cladogram shown in Fig.~6(a).  The problem is that these contractions lose 
some of the information encoded in the initial Hennig tree.  For instance, the structure of $Q$ expresses the hypothesis 
that $B$ is the common ancestor of $E$ and $F$, but these three vertices become mutually indistinguishable upon 
contraction. Instead, one should interpret each hollow circle as the collection of characters that distinguishes its 
child species from its ancestors.  For instance, the vertex whose only child is $E$ represents the characters fixed in 
$E$, but not fixed in the ancestor $B$ (nor in any other species in the table).  Therefore, the cladogram depicted in 
Fig.~6(b) contains the same information as the original Hennig tree $H$.  (This should really be interpreted as a 
partial cladogram; that is, we do not claim that we obtain a complete list of characters in this way.)  A possible 
biological interpretation of this cladogram is that the sampled species $B$ is regarded not as an ancestor of $E$, but 
as 
a species indistinguishable by character analysis from the common ancestor of $E$ and $F$.  On the other hand, it is 
easier to represent the phylogenetic relationships between the sampled taxa by a Hennig tree (or, equivalently, the 
phylogenetic tree to which it is equivalent via Algorithms A and B), which encodes relationships more simply and 
efficiently. It is less clear how to pass from an arbitrary cladogram (in which internal vertices can have zero, or more 
than one two, adjacent leaves) to a Hennig tree (see, e.g., Cracraft, 1974; Harper, 1976; Platnick, 1977; Wiley, 1979a, 
b, and 1981a). To do so would seem to require a biological interpretation of edges and nodes.

\section*{Apomorphy-Based Monophyly}

One is very rarely able to recognize a sampled specimen as representing an ancestral species. What we observe are 
specimens and their properties. Hennig's (1966:79-80) attitude was that characters do not make the taxon; the ancestor 
makes the taxon and characters are tools for hypothesizing common ancestry. Thus, synapomorphies are the historical 
marks of common ancestry, or more properly, assertions of the presence of common ancestors in the history of descent. 
Relative to circumscription, we can say that all monophyletic groups are circumscribed by synapomorphies or they are 
circumscribed by no testable properties. However, the third manner of circumscribing clades according to the Phylocode 
is incorrect: ``An apomorphy-based clade is a clade originating from the ancestor in which a particular derived 
character state (apomorphy) originated'' (Phylocode Note 2.1.4). In fact, rarely can we assert that we know when a 
character originated, only when it was hypothesized to be ``fixed'' in a particular lineage or when it remains variable 
among the taxa studied.

\section*{Acknowledgements}

EOW thanks David Hull (Northwestern University) for sending a copy of a manuscript Hull never published entitled 
``Hierarchies and Hierarchies'' that touched upon the problems associated with process/pattern and tree/cladogram 
controversies; and for what must have seemed to him hours of discussion on things phylogenetic and philosophical 
regarding the subject. We also thank Shannon DeVaney (University of Kansas) for reading the manuscript and providing a 
critical review.

\section*{Literature Cited}

Camin, J. H., and R. R. Sokal.  1965.  A method for deducing branching sequences in phylogeny.  Evolution 19: 311-326.

Cracraft, J. 1974. Phylogenetic models and classification. Syst. Zool. 23:71-90.

de Queiroz, K. 2007. Toward and integrated system of clade names. Syst. Biol. 56:956-974.

de Queiroz, K. and J. Gauthier. 1990. Phylogeny as a central principle in taxonomy: Phylogenetic definitions of taxon 
names. Syst. Zool. 39:307-322.

de Queiroz, K., and J. Gauthier. 1992. Phylogenetic taxonomy. Annu. Rev. Ecol. Syst. 23:449-480.

de Queiroz, K., and J. Gauthier. 1994. Toward a phylogenetic system of biological nomenclature. Trends Ecol. Evol. 
9:27-31.

Gregg, J. R. 1954. The Language of taxonomy. An application of symbolic logic to the study of classificatory systems. 
Columbia University Press, New York.

Harper, C. W., Jr. 1979. Phylogenetic inference in paleontology. J. Paleontol. 50:180-193.

Hendy, M. D., and D. Penny. 1989. A framework for the quantitative study of evolutionary trees. Syst. Zool. 38:297.309.

Hennig, W. 1966. Phylogenetic systematics. University of Illinois Press, Urbana.

Hull, D. L. 1979. The limits of cladism. Syst. Zool. 28:416-440

Nelson, G.  1979 ms. Cladograms and trees. (Widely circulated but never published.)

Nelson, G., and N. I. Platnick. 1981. Systematics and Biogeography: Cladistics and Vicariance. Columbia University 
press, New York.

Platnick, N. I. 1977. Cladograms, phylogenetic trees, and hypothesis testing. Syst. Zool. 26:438-442.

West, D. B. 2006.  Introduction to graph theory, 2nd edn.  Prentice Hall, Upper Saddle River, NJ.

Wiens, J. J., and M. R. Servedio. 2000. Species delimitation in systematics: inferring diagnostic differences between 
species. Proc. Royal Soc. of London, Series B 267:631.636.

Wiley, E. O. 1979a. Cladograms and phylogenetic trees. Syst. Zool. 28:88-92.

Wiley, E. O. 1979b. Ancestors, species, and cladograms.-Remarks on the symposium. Pages 211-225 in Phylogenetic Analysis 
and paleontology (J. Cracraft and N. Eldredge, eds.). Columbia University Press, New York.

Wiley, E. O. 1981. Phylogenetics. The theory and practice of phylogenetic systematics. Wiley-Interscience, New York.

\includegraphics{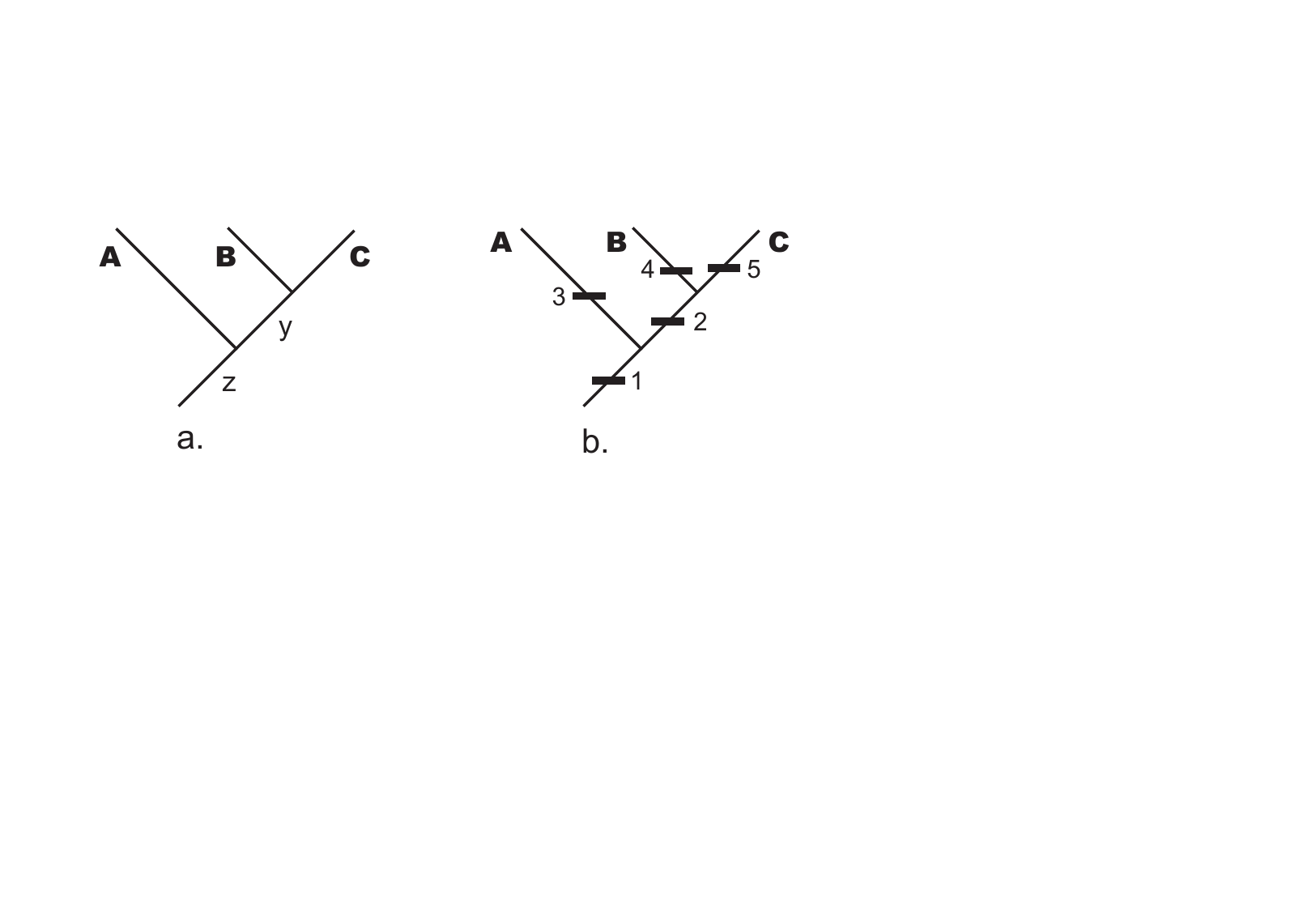}
Figure 1. (a) An example of a phylogenetic tree, indicating the evolutionary relationship among the sampled 
taxa A, B, C and their unsampled ancestral species y and z. (b) The same tree with character data shown.
The names of the internal 
edges have been omitted for clarity.  In each case, taxon names are displaced from the leaf position to emphasize that 
the edge is the taxon.
\pagebreak

\resizebox{6in}{4in}{\includegraphics{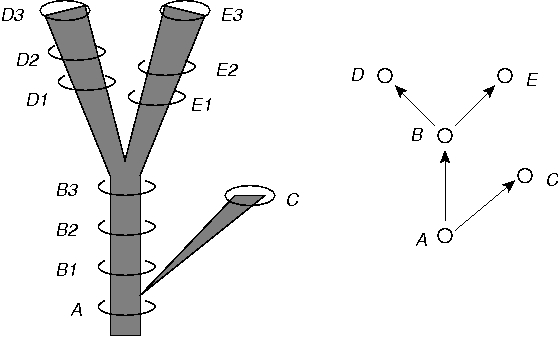}}
\vfill
Figure 2. Modified version of Figure 14 of Hennig (1966:59) entitled ``The species category in the time dimension''.
Left: a phylogenetic tree. Letters are symbols for species and the number applied to the numbers are labels for samples
of each species considered at a particular time period. Right: a Hennig tree with single-headed arrows symbolizing
relationship statements and circles representing species. Note the correspondence between the lineages on the left and
the circles on the right, as shown by the brackets and double-headed arrows for selected lineages and vertices. Redrawn
from Hennig (1966).
\vfill
\pagebreak

\resizebox{6in}{3.75in}{\includegraphics{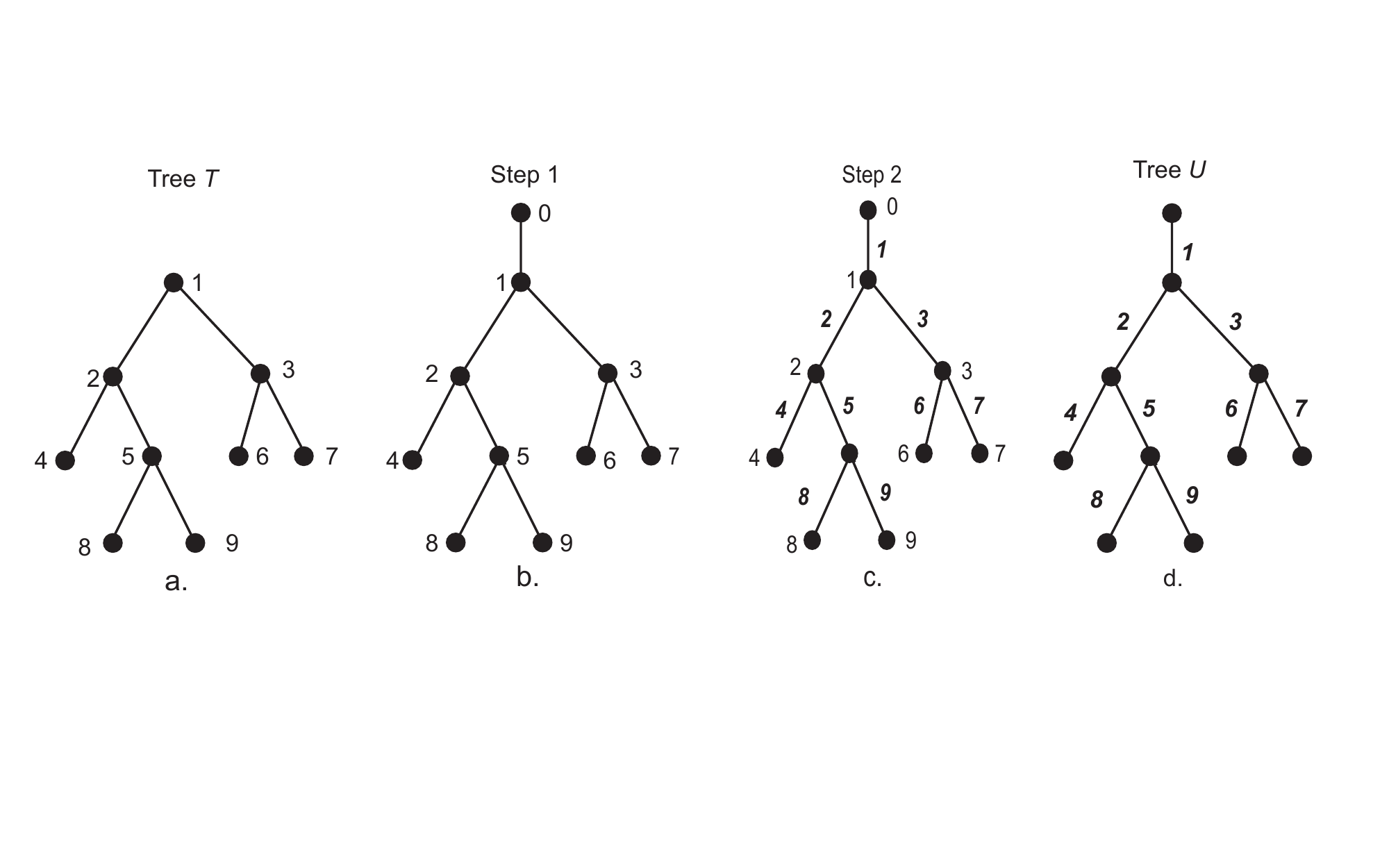}}
Figure~3(a-d). The steps of Algorithm~A, read left to right.  Reading right to left illustrates Algorithm~B.
\pagebreak

\resizebox{6in}{3.75in}{\includegraphics{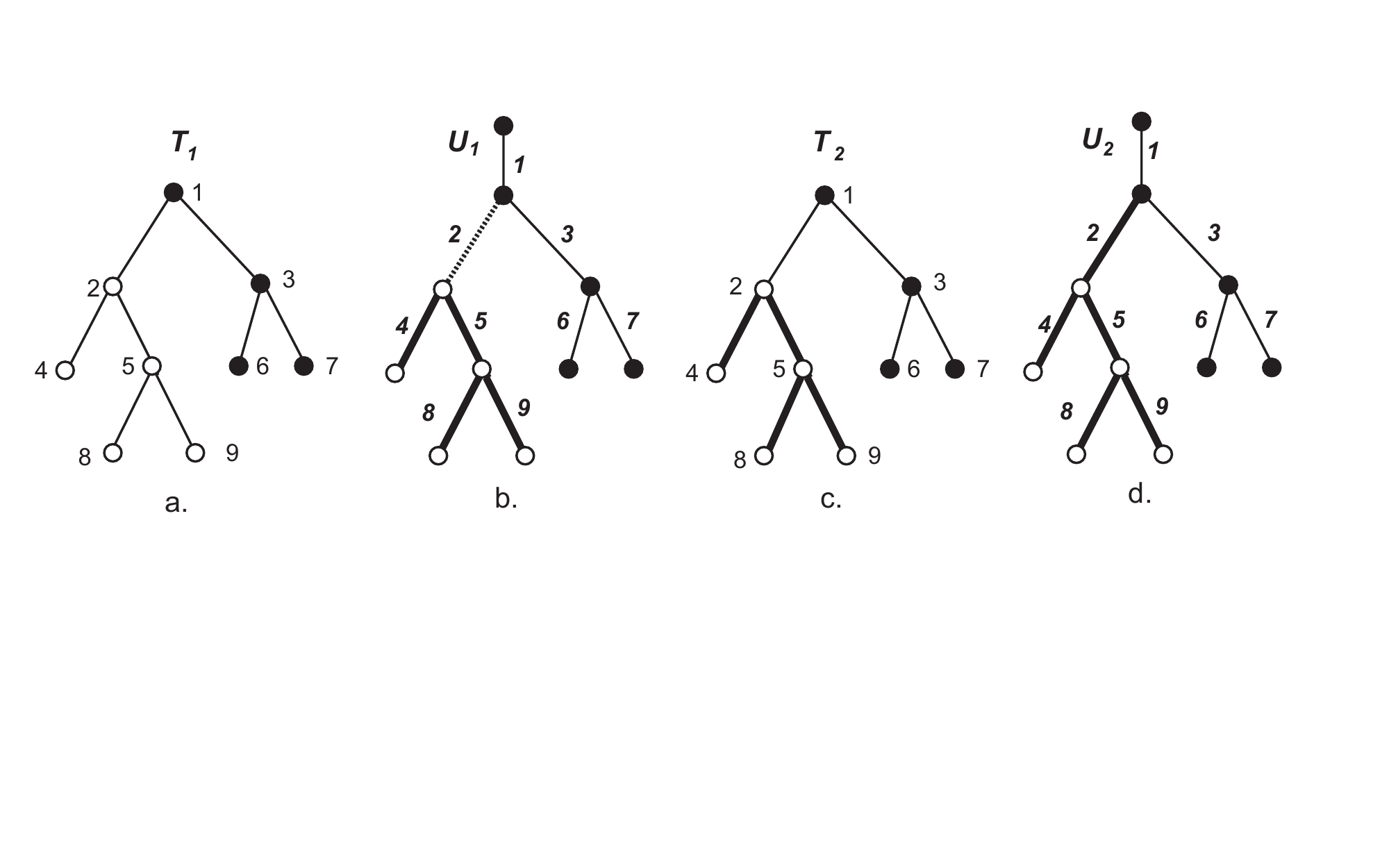}}
Figure 4. (a) A Hennig tree $T$. The vertices 4, 5, 8, 9 do not form a subtree, even though edges 4, 5, 8, 9 form a
subtree of the corresponding phylogenetic tree $U$ shown in (b). In contrast, the subtree of $T$ formed by vertices 2,
4, 5, 8, 9, as shown in (c), corresponds to the planted subtree of $U$ shown in (d). The figure also illustrates
possible circumscriptions of the terminal taxa 4, 8, 9. Heavy lines denote edges included in the classification.
(Mis)applying node-based circumscription to $U$ results in the polyphyletic group 4, 5, 8, 9; as shown in (b), there is
no edge connection to the sister group comprising the terminals 6 and 7 because 2 remains logically unclassified (dashed
line).  In contrast, a node-based circumscription of $T$ or a stem-based circumscription of $U$ (shown in (c) and (d))
yields the monophyletic group composed of the terminal taxa 4, 8, and 9 and their inferred ancestors 2 and 5.
\pagebreak

\includegraphics{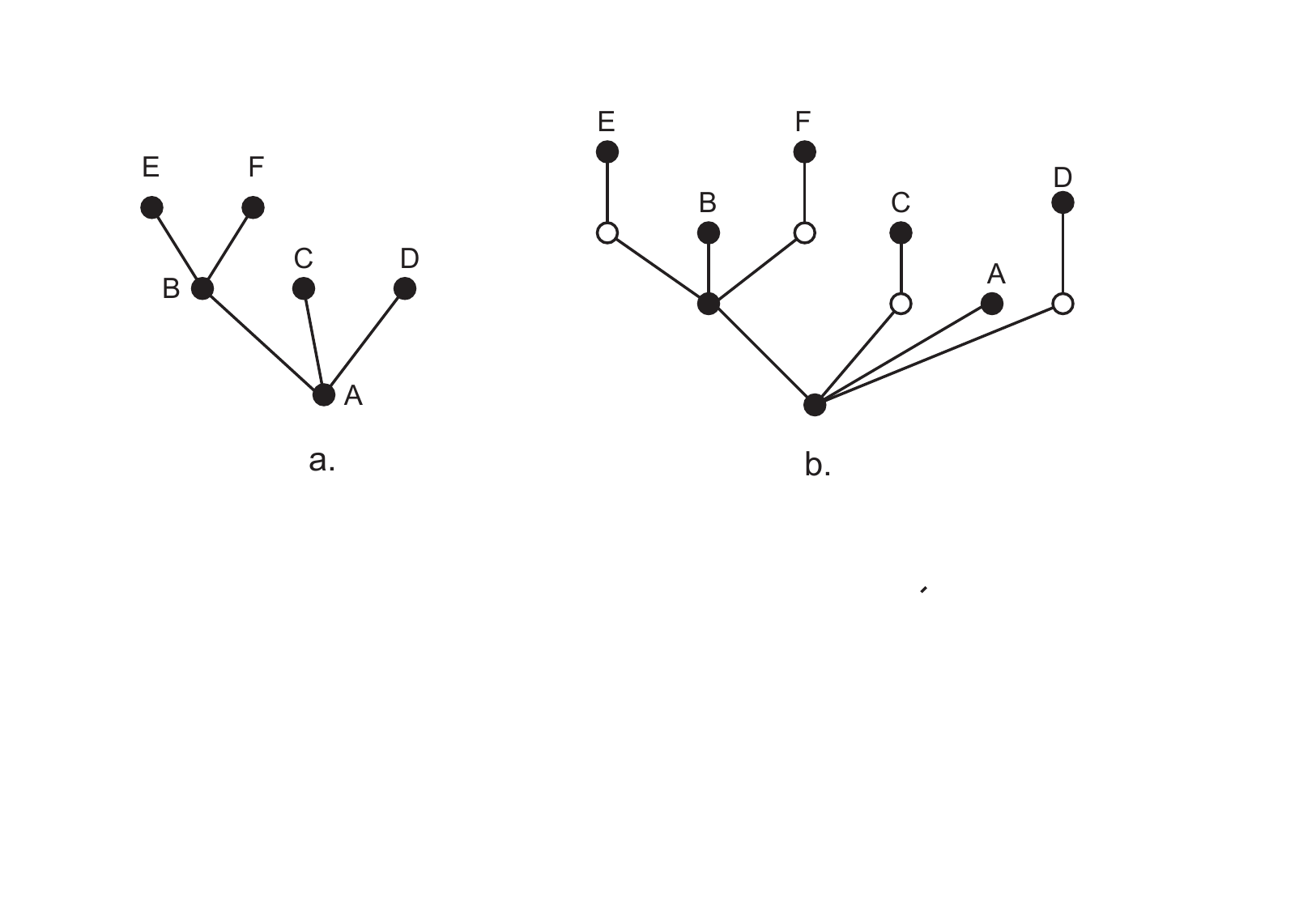}
Figure 5. (a)  A Hennig tree $H$.  (b)  The ``quasi-cladogram'' $Q$ produced from $H$ by Algorithm~C, and from which
$H$ can be recovered by Algorithm~D.
\pagebreak

\includegraphics{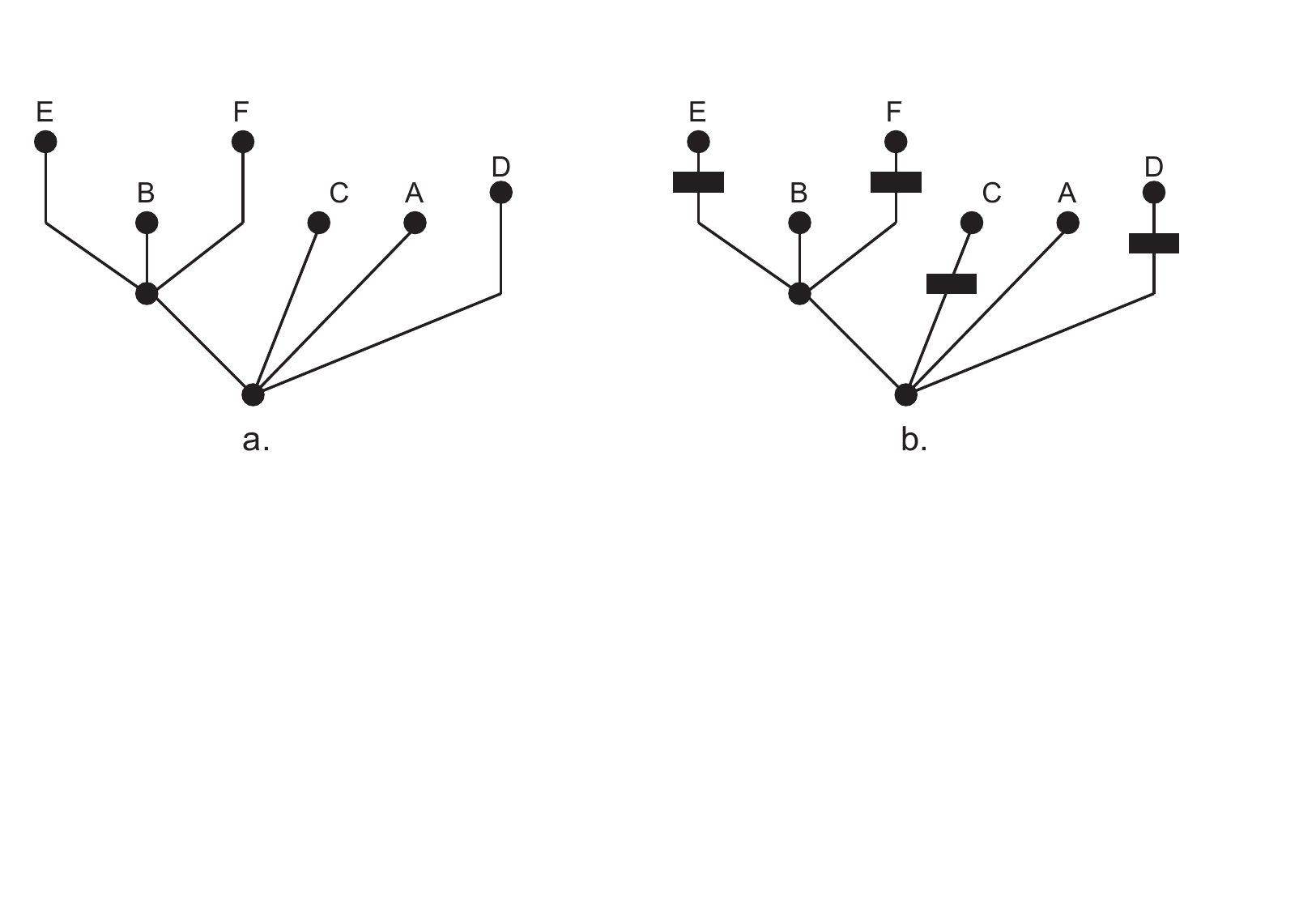}
Figure 6. A cladogram incorrectly obtained from $Q$ (see Fig.~5(b)) by contraction.  Note that vertices B, E, F are
now indistinguishable, as are A, C, D.  (b) The correct cladogram corresponding to $Q$.

\vfill

\end{document}